\documentclass[12pt]{article}
\usepackage{amssymb,amsmath,epsfig}

\begin{document}
\title{\bf Thermodynamics in $f(R,T)$ Theory of Gravity}

\author{M. Sharif \thanks{msharif.math@pu.edu.pk} and M. Zubair
\thanks{mzubairkk@gmail.com}\\\\
Department of Mathematics, University of the Punjab,\\
Quaid-e-Azam Campus, Lahore-54590, Pakistan.}

\date{}

\maketitle

\begin{abstract}
A non-equilibrium picture of thermodynamics is discussed at the
apparent horizon of FRW universe in $f(R,T)$ gravity, where $R$ is
the Ricci scalar and $T$ is the trace of the energy-momentum tensor.
We take two forms of the energy-momentum tensor of dark components
and demonstrate that equilibrium description of thermodynamics is
not achievable in both cases. We check the validity of the first and
second law of thermodynamics in this scenario. It is shown that the
Friedmann equations can be expressed in the form of first law of
thermodynamics $T_hdS'_h+T_hd_{\jmath}S'=-dE'+W'dV$, where
$d_{\jmath}S'$ is the entropy production term. Finally, we conclude
that the second law of thermodynamics holds both in phantom and
non-phantom phases.
\end{abstract}

\section{Introduction}

Cosmic observations from anisotropy of the Cosmic Microwave
Background (CMB) \cite{1}, supernova type Ia (SNeIa) \cite{2}, large
scale structure \cite{3}, baryon acoustic oscillations \cite{4} and
weak lensing \cite{5} indicate that expansion of the universe is
speeding up rather than decelerating. The present accelerated
expansion is driven by gravitationally repulsive dominant energy
component known as \emph{dark energy} (DE). There are two
representative directions to address the issue of cosmic
acceleration. One is to introduce the "exotic energy component" in
the context of General Relativity (GR). Several candidates have been
proposed \cite{6}-\cite{10} in this perspective to explore the
nature of DE. The other direction is to modify the Einstein
Lagrangian i.e., modified gravity theory such as $f(R)$ gravity
\cite{11}.

The discovery of black hole thermodynamics set up a significant
connection between gravity and thermodynamics \cite{12}. The Hawking
temperature $T=\frac{|\kappa_{sg}|}{2\pi}$, where $\kappa_{sg}$ is
the surface gravity, and horizon entropy $S=\frac{A}{4G}$ obey the
first law of thermodynamics \cite{12}-\cite{14}. Jacobson \cite{15}
showed that it is indeed possible to derive the Einstein field
equations in Rindler spacetime by using the Clausius relation
$TdS={\delta}Q$ and proportionality of entropy to the horizon area.
Here ${\delta}Q$ and $T$ are the energy flux across the horizon and
Unruh temperature respectively, viewed by an accelerated observer
just inside the horizon. Frolov and Kofman \cite{16} employed this
approach to quasi-de Sitter geometry of inflationary universe with
the relation $-dE=TdS$ to calculate energy flux of slowly rolling
background scalar field.

Cai and Kim \cite{17} derived the Freidmann equations of the FRW
universe with any spatial curvature from the first law of
thermodynamics for the entropy of the apparent horizon. Later, Akbar
and Cai \cite{18} showed that the Friedmann equations in GR can be
written in the form $dE=TdS+WdV$ at the apparent horizon, where
$E={\rho}V$ is the total energy inside the apparent horizon and
$W=\frac{1}{2}(\rho-p)$ is the work density. The connection between
gravity and thermodynamics has been revealed in modified theories of
gravity including Gauss-Bonnet gravity \cite{18}, Lovelock gravity
\cite{19,20}, braneworld gravity \cite{21}, non-linear gravity
\cite{22}-\cite{27} and scalar-tensor gravity \cite{19,28,29}. In
$f(R)$ gravity and scalar-tensor theory, non-equilibrium description
of thermodynamics is required \cite{22}-\cite{29} so that the
Clausius relation is modified in the form $TdS={\delta}Q+d\bar{S}$.
Here, $d\bar{S}$ is the additional entropy production term.

Recently, Harko et al. \cite{30} generalized $f(R)$ gravity by
introducing an arbitrary function of the Ricci scalar $R$ and the
trace of the energy-momentum tensor $T$. The dependence of $T$ may
be introduced by exotic imperfect fluids or quantum effects
(conformal anomaly). As a result of coupling between matter and
geometry motion of test particles is nongeodesic and an extra
acceleration is always present. In $f(R,T)$ gravity, cosmic
acceleration may result not only due to geometrical contribution to
the total cosmic energy density but it also depends on matter
contents. This theory can be applied to explore several issues of
current interest and may lead to some major differences. Houndjo
\cite{31} developed the cosmological reconstruction of $f(R,T)$
gravity for $f(R,T)=f_1(R)+f_2(T)$ and discussed transition of
matter dominated phase to an acceleration phase.

In a recent paper \cite{26}, Bamba and Geng investigated laws of
thermodynamics in $f(R)$ gravity. It is argued that equilibrium
description exists in $f(R)$ gravity. The equilibrium description of
thermodynamics in modified gravitational theories is still under
debate as various alternative treatments \cite{31a} have been
proposed to reinterpret the non-equilibrium picture. These recent
studies have motivated us to explore whether the equilibrium
description can be obtained in the framework of $f(R,T)$ gravity.
The study of connection between gravity and thermodynamics in
$f(R,T)$ may provide some specific results which would discriminate
this theory from various theories of modified gravity.

In this paper, we examine whether an equilibrium description of
thermodynamics is possible in such a modified theory of gravity. The
horizon entropy is constructed from the first law of thermodynamics
corresponding to the Friedmann equations. We explore the generalized
second law of thermodynamics (GSLT) and find out the necessary
condition for its validity. The paper is organized as follows: In
the next section, we review $f(R,T)$ gravity and formulate the field
equations of FRW universe. Section \textbf{3} investigates the first
and second laws of thermodynamics. In section \textbf{4}, the
Friedmann equations are reformulated by redefining the dark
components to explore the possible change in thermodynamics.
Finally, section \textbf{5} is devoted to the concluding remarks.

\section{$f(R,T)$ Gravity}

The action of $f(R,T)$ theory of gravity is given by \cite{30}
\begin{equation}\label{1}
\mathcal{A}=\int{dx^4\sqrt{-g}\left[\frac{f(R,T)}{16{\pi}G}+\mathcal{L}_{(matter)}\right]},
\end{equation}
where $\mathcal{L}_{(matter)}$ determines matter contents of the
universe. The energy-momentum tensor of matter is defined as
\cite{32}
\begin{equation}\label{2}
T_{{\mu}{\nu}}^{(matter)}=-\frac{2}{\sqrt{-g}}\frac{\delta(\sqrt{-g}
{\mathcal{\mathcal{L}}_{(matter)}})}{\delta{g^{{\mu}{\nu}}}}.
\end{equation}
We assume that the matter Lagrangian density depends only on the
metric tensor components $g_{{\mu}{\nu}}$ so that
\begin{equation}\label{3}
T_{{\mu}{\nu}}^{(matter)}=g_{{\mu}{\nu}}\mathcal{L}_{(matter)}-\frac{2{\partial}
{\mathcal{L}_{(matter)}}}{\partial{g^{{\mu}{\nu}}}}.
\end{equation}
Variation of the action (\ref{1}) with respect to the metric tensor
yields the field equations of $f(R,T)$ gravity as
\begin{eqnarray}\label{4}
&&R_{{\mu}{\nu}}f_{R}(R,T)-\frac{1}{2}g_{{\mu}{\nu}}f(R,T)+(g_{{\mu}{\nu}}
{\Box}-{\nabla}_{\mu}{\nabla}_{\nu})f_{R}(R,T)\nonumber\\&=&8{\pi}G
T_{{\mu}{\nu}}^{(matter)}-f_{T}(R,T)T_{{\mu}{\nu}}^{(matter)}-f_{T}(R,T)\Theta_{{\mu}{\nu}},
\end{eqnarray}
where ${\nabla}_{\mu}$ is the covariant derivative associated with
the Levi-Civita connection of the metric and
${\Box}={\nabla}_{\mu}{\nabla}^{\mu}$. We denote
$f_{R}(R,T)={\partial}f(R,T)/{\partial}R$,
$f_{T}(R,T)={\partial}f(R,T)/{\partial}T$ and
$\Theta_{{\mu}{\nu}}=\frac{g^{\alpha{\beta}}{\delta}T_{{\alpha}{\beta}}}{{\delta}
g^{\mu{\nu}}}.$ The choice of $f(R,T){\equiv}f(R)$ results in the
field equations of $f(R)$ gravity.

The energy-momentum tensor of matter is defined as
\begin{equation}\label{5}
T_{{\mu}{\nu}}^{(matter)}=({\rho}_m+p_m)u_{\mu}u_{\nu}+p_{m}g_{{\mu}{\nu}},
\end{equation}
where $u_{\mu}$ is the four velocity of the fluid. If we take
$\mathcal{L}_{(matter)}=-p_m$, then $\Theta_{{\mu}{\nu}}$ becomes
\begin{equation}\label{6}
\Theta_{{\mu}{\nu}}=-2T_{{\mu}{\nu}}^{(matter)}-p_{m}g_{{\mu}{\nu}}.
\end{equation}
Consequently, the field equations (\ref{4}) lead to
\begin{eqnarray}\label{7}
&&R_{{\mu}{\nu}}f_{R}(R,T)-\frac{1}{2}g_{{\mu}{\nu}}f(R,T)+(g_{{\mu}{\nu}}
{\Box}-{\nabla}_{\mu}{\nabla}_{\nu})f_{R}(R,T)\nonumber\\&=&8{\pi}G
T_{{\mu}{\nu}}^{(matter)}+T_{{\mu}{\nu}}^{(matter)}f_{T}(R,T)+p_{m}g_{{\mu}{\nu}}f_{T}(R,T).
\end{eqnarray}
The field equations in $f(R,T)$ gravity depend on a source term,
representing the variation of the energy-momentum tensor of matter
with respect to the metric. We consider only the non-relativistic
matter (cold \emph{dark} matter and baryons) with $p_{m}=0$,
therefore the contribution of $T$ comes only from ordinary matters.
Thus, Eq.(\ref{7}) can be written as an effective Einstein field
equation of the form
\begin{equation}\label{8}
R_{{\mu}{\nu}}-\frac{1}{2}Rg_{{\mu}{\nu}}
=8{\pi}G_{eff}T_{{\mu}{\nu}}^{(matter)}+{T'}_{{\mu}{\nu}}^{(d)},
\end{equation}
where
\begin{equation*}
G_{eff}=\frac{1}{f_{R}(R,T)}\left(G+\frac{f_{T}(R,T)}{8\pi}\right)
\end{equation*}
is the effective gravitational matter dependent coupling in $f(R,T)$
gravity and
\begin{equation}\label{9}
{T'}_{{\mu}{\nu}}^{(d)}=\frac{1}{f_{R}(R,T)}\left[\frac{1}{2}g_{\mu\nu}
(f(R,T)-Rf_{R}(R,T))+({\nabla}_{\mu}{\nabla}_{\nu}-g_{{\mu}{\nu}}{\Box})
f_{R}(R,T)\right]
\end{equation}
is the energy-momentum tensor of \emph{dark components}. Here, prime
means non-equilibrium description of the field equations.

The FRW universe is described by the metric
\begin{equation}\label{10}
ds^{2}=h_{\alpha\beta}dx^{\alpha}dx^{\beta}+\tilde{r}^{2}d{\Omega}^2,
\end{equation}
where $\tilde{r}=a(t)r$ and $x^{0}=t,~x^{1}=r$ with the
2-dimensional metric $h_{\alpha{\beta}}=diag(-1,a^2/(1-kr^2))$. Here
$a(t)$ is the scale factor, $k$ is the cosmic curvature and
$d{\Omega}^2$ is the metric of 2-dimensional sphere with unit
radius. In FRW background, the gravitational field equations are
given by
\begin{eqnarray}\label{11}
3\left(H^2+\frac{k}{a^2}\right)&=&8{\pi}G_{eff}{\rho}_m+\frac{1}{f_{R}}
\left[\frac{1}{2}(Rf_{R}-f)-3H(\dot{R}f_{RR}\right.\\\nonumber&+&\left.
\dot{T}f_{RT})\right],\\\nonumber-\left(2\dot{H}+3H^2+\frac{k}{a^2}\right)
&=&\frac{1}{f_{R}}\left[-\frac{1}{2}(Rf_{R}-f)+2H(\dot{R}f_{RR}+\dot{T}
f_{RT})+\ddot{R}f_{RR}\right.\\\label{12}&+&\left.
\dot{R}^2f_{RRR}+2\dot{R}\dot{T}f_{RRT}
+\ddot{T}f_{RT}+\dot{T}^2f_{RTT}\right].
\end{eqnarray}
These can be rewritten as
\begin{eqnarray}\label{13}
3\left(H^2+\frac{k}{a^2}\right)&=&8{\pi}G_{eff}({\rho}_m+{\rho}'_d),
\\\label{14}-2\left(\dot{H}-\frac{k}{a^2}\right)&=&8{\pi}G_{eff}({\rho}_m
+{\rho}'_d+{p}'_d),
\end{eqnarray}
where ${\rho}'_d$ and ${p}'_d$ are the energy density and pressure
of \emph{dark components}
\begin{eqnarray}\label{15}
{\rho}'_d&=&\frac{1}{8{\pi}G\mathcal{F}}\left[\frac{1}{2}(Rf_{R}-f)-3H(\dot{R}f_{RR}
+\dot{T}f_{RT})\right],\\\nonumber{p}'_d&=&\frac{1}{8{\pi}G\mathcal{F}}\left[-\frac{1}{2}
(Rf_{R}-f)+2H(\dot{R}f_{RR}+\dot{T}f_{RT})+\ddot{R}f_{RR}+\dot{R}^2f_{RRR}\right.\\\label{16}&+&\left.
2\dot{R}\dot{T}f_{RRT}+\ddot{T}f_{RT}+\dot{T}^2f_{RTT}\right].
\end{eqnarray}

Here $\mathcal{F}=1+\frac{f_{T}(R,T)}{8{\pi}G}$. The equation of
state (EoS) parameter of \emph{dark} fluid ${\omega}'_d$ is obtained
as ($p'_d=\omega'_d\rho'_d$)
\begin{equation}\label{16a}
{\omega}'_d=-1+\frac{\ddot{R}f_{RR}+\dot{R}^2f_{RRR}+2\dot{R}\dot{T}f_{RRT}
+\ddot{T}f_{RT}+\dot{T}^2f_{RTT}-H(\dot{R}f_{RR}+\dot{T}f_{RT})}
{\frac{1}{2}(Rf_{R}-f)-3H(\dot{R}f_{RR}+\dot{T}f_{RT})}.
\end{equation}

The semi-conservation equation of ordinary matter is given by
\begin{eqnarray}\label{16b}
\dot{\rho}+3H\rho=q.
\end{eqnarray}
The energy-momentum tensor of \emph{dark components} may satisfy the
similar conservation laws
\begin{eqnarray}\label{16c}
\dot{\rho}_d+3H(\rho_d+p_d)&=&q_d,\\\label{16d}
\dot{\rho}_{tot}+3H(\rho_{tot}+p_{tot})&=&q_{tot},
\end{eqnarray}
where $\rho_{tot}=\rho_m+\rho_d,~p_{tot}=p_d$ and $q_{tot}=q+q_d$ is
the total energy exchange term and $q_d$ is the energy exchange term
of dark components. Substituting Eqs.(\ref{13}) and (\ref{14}) in
the above equation, we obtain
\begin{eqnarray}\label{16e}
q_{tot}=\frac{3}{8{\pi}G}(H^2+\frac{k}{a^2})\partial_t
\left(\frac{f_{R}}{\mathcal{F}}\right).
\end{eqnarray}
The relation of energy exchange term in $f(R)$ gravity can be
recovered if $\mathcal{F}=1$. In GR, $q_{tot}=0$ for the choice
$f(R,T)=R$.

\section{Laws of Thermodynamics}

In this section, we examine the validity of the first and second law
of thermodynamics in $f(R,T)$ gravity for FRW universe.

\subsection{First Law of Thermodynamics}

Here we investigate the validity of the first law of thermodynamics
in $f(R,T)$ gravity at the apparent horizon of FRW universe. The
dynamical apparent horizon is determined by the relation
$h^{\alpha\beta}\partial_{\alpha}\tilde{r}\partial_{\beta}\tilde{r}=0$
which leads to the radius of apparent horizon for FRW universe
\begin{equation}\label{17}
\tilde{r}_A=\frac{1}{\sqrt{H^2+\frac{k}{a^2}}}.
\end{equation}
The associated temperature of the apparent horizon is defined
through the surface gravity $\kappa_{sg}$ as
\begin{equation}\label{18}
T_h=\frac{|\kappa_{sg}|}{2\pi},
\end{equation}
where $\kappa_{sg}$ is given by \cite{17}
\begin{eqnarray}\label{19}
\kappa_{sg}&=&\frac{1}{2\sqrt{-h}}\partial_{\alpha}(\sqrt{-h}h^
{\alpha\beta}\partial_{\beta}\tilde{r}_A)=-\frac{1}{\tilde{r}_A}
(1-\frac{\dot{\tilde{r}}_A}{2H\tilde{r}_A})\nonumber\\
&=&-\frac{\tilde{r}_A}{2}(2H^2+\dot{H}+\frac{k}{a^2}).
\end{eqnarray}

In GR, the horizon entropy is given by the Bekenstein-Hawking
relation $S_h=A/4G$, where $A=4{\pi}\tilde{r}^2_A$ is the area of
the apparent horizon \cite{12}-\cite{14}. In the context of modified
gravitational theories, Wald \cite{33} proposed that entropy of
black hole solutions with bifurcate Killing horizons is a Noether
charge entropy. It depends on the variation of Lagrangian density of
modified gravitational theories with respect to Riemann tensor. Wald
entropy is equal to quarter of horizon area in units of effective
gravitational coupling i.e, ${S}'_h=A/4G_{eff}$ \cite{33a}. In
$f(R,T)$ gravity, the Wald entropy is expressed as
\begin{equation}\label{20}
{S}'_h=\frac{Af_{R}}{4G\mathcal{F}}.
\end{equation}
Taking the time derivative of Eq.(\ref{17}) and using (\ref{14}), it
follows that
\begin{equation}\label{21}
f_Rd\tilde{r}_A=4{\pi}G\tilde{r}^3_A({\rho}'_{tot}+{p}'_{tot})H\mathcal{F}dt,
\end{equation}
where ${\rho}'_{tot}={\rho}'_m+{\rho}'_d$ and ${p}'_{tot}={p}'_d$.
$d\tilde{r}_A$ is the infinitesimal change in radius of the apparent
horizon during a time interval $dt$. Using Eqs.(\ref{20}) and
(\ref{21}), we obtain
\begin{equation}\label{22}
\frac{1}{2{\pi}\tilde{r}_A}d{S}'_h=4{\pi}\tilde{r}^3_A({\rho}'_{tot}
+{p}'_{tot})Hdt+\frac{\tilde{r}_A}{2G\mathcal{F}}df_R+\frac{\tilde{r}_A
f_R}{2G}d\left(\frac{1}{\mathcal{F}}\right).
\end{equation}
If we multiply both sides of this equation with a factor
$(1-\dot{\tilde{r}}_A/2H\tilde{r}_A)$, it follows that
\begin{eqnarray}\label{23}
T_hd{S}'_h&=&4{\pi}\tilde{r}^3_A({\rho}'_{tot}+{p}'_{tot})
Hdt-2{\pi}\tilde{r}^2_A({\rho}'_{tot}+{p}'_{tot})d\tilde{r}_A+\frac{{\pi}
\tilde{r}^2_AT_hdf_R}{G\mathcal{F}}\nonumber\\&+&\frac{{\pi}\tilde{r}^2_AT_hf_R}{G}
d\left(\frac{1}{\mathcal{F}}\right).
\end{eqnarray}

Now, we define energy of the universe within the apparent horizon.
The Misner-Sharp energy \cite{34} is defined as
$E=\frac{\tilde{r}_A}{2G}$ which can be written in $f(R,T)$ gravity
as \cite{24,28}
\begin{equation}\label{24}
E'=\frac{\tilde{r}_A}{2G_{eff}}.
\end{equation}
In terms of volume $V=\frac{4}{3}{\pi}\tilde{r}^3_A$, we obtain
\begin{equation}\label{25}
E'=\frac{3V}{8{\pi}G_{eff}}\left(H^2+\frac{k}{a^2}\right)
=V{\rho}'_{tot}
\end{equation}
which represents the total energy inside the sphere of radius
$\tilde{r}_A$. It is obvious that $E'>0$, if
$G_{eff}=\frac{G\mathcal{F}}{f_R}>0$ so that the effective
gravitational coupling constant in $f(R,T)$ gravity should be
positive. It follows from Eqs.(\ref{13}) and (\ref{25}) that
\begin{equation}\label{26}
d{E}'=-4{\pi}\tilde{r}^3_A({\rho}'_{tot}+{p}'_{tot})Hdt+4{\pi}
\tilde{r}^2_A{\rho}'_{tot}d\tilde{r}_A+\frac{\tilde{r}_Adf_R}{2G\mathcal{F}}+
\frac{\tilde{r}_Af_R}{2G}d\left(\frac{1}{\mathcal{F}}\right).
\end{equation}
Using Eq.(\ref{26}) in (\ref{23}), it follows that
\begin{equation}\label{27}
T_hd{S}'_h=-d{E}'+{W}'dV+\frac
{(1+2{\pi}\tilde{r}_AT_h)\tilde{r}_Adf_R}{2G\mathcal{F}}+\frac{
(1+2{\pi}\tilde{r}_AT_h)\tilde{r}_Af_R}{2G}d\left(\frac{1}
{\mathcal{F}}\right),
\end{equation}
which includes the work density
${W}'=\frac{1}{2}({\rho}'_{tot}-{p}'_{tot})$ \cite{35}. This can be
rewritten as
\begin{equation}\label{28}
T_hd{S}'_h+T_hd_{\jmath}{S}'_h=-d{E}'+{W}'dV,
\end{equation}
where
\begin{equation}\label{29}
d_{\jmath}{S}'_h=-\frac{\tilde{r}_A}{2GT_h}(1+2{\pi}\tilde{r}_AT_h)
d\left(\frac{f_R}{\mathcal{F}}\right)=-\frac{\mathcal{F}
(E'+S'_hT_h)}{T_hf_R} d\left(\frac{f_R}{\mathcal{F}}\right).
\end{equation}
When we compare the cosmological setup of $f(R,T)$ gravity with GR,
Gauss-Bonnet gravity and Lovelock gravity \cite{18}-\cite{20}, we
obtain an auxiliary term in the first law of thermodynamics. This
additional term $d_{\jmath}S'_h$ may be interpreted as entropy
production term developed due to the non-equilibrium framework in
$f(R,T)$ gravity. This result corresponds to the first law of
thermodynamics in non-equilibrium description of $f(R)$ gravity
\cite{26} for $f(R,T)=f(R)$. If we assume $f(R,T)=R$, then the
traditional first law of thermodynamics in GR can be achieved.

\subsection{Generalized Second Law of Thermodynamics}

Recently, the GSLT has been studied in the context of modified
gravitational theories \cite{25}-\cite{28}. It may be interesting to
investigate its validity in $f(R,T)$ gravity. For this purpose, we
have to show that \cite{28}
\begin{equation}\label{30}
\dot{S}'_h+d_\jmath\dot{S}'_h+\dot{S}'_{tot}\geq0,
\end{equation}
where ${S}'_h$ is the horizon entropy in $f(R,T)$ gravity,
$d_\jmath\dot{S}'_h=\partial_t(d_\jmath{S}'_h)$ and $S'_{tot}$ is
the entropy due to all the matter and energy sources inside the
horizon. The Gibb's equation including all matter and energy fluid
is given by \cite{36}
\begin{equation}\label{31}
T_{tot}dS'_{tot}=d({\rho}'_{tot}V)+{p}'_{tot}dV,
\end{equation}
where $T_{tot}$ is the temperature of total energy inside the
horizon. We assume that $T_{tot}$ is proportional to the temperature
of apparent horizon \cite{25,28}, i.e., $T_{tot}=bT_h$, where
$0<b<1$ to ensure that temperature being positive and smaller than
the horizon temperature.

Substituting Eqs.(\ref{28}) and (\ref{31}) in Eq.(\ref{30}), we
obtain
\begin{equation}\label{32}
\dot{S}'_h+d_\jmath\dot{S}'_h+\dot{S}'_{tot}=
\frac{24\pi{\Xi}}{\tilde{r}_AbR}\geq0,
\end{equation}
where
\begin{equation*}
\Xi=(1-b)\dot{\rho}'_{tot}V+(1-\frac{b}{2})(\rho'_{tot}+p'_{tot})\dot{V}
\end{equation*}
is the universal condition to protect the GSLT in modified
gravitational theories \cite{28}. Using Eqs.(\ref{13}) and
(\ref{14}), condition (\ref{30}) is reduced to
\begin{equation}\label{33}
\frac{12\pi\mathcal{X}}{bRG\mathcal{F}(H^2+\frac{k}{a^2})^2}\geq0,
\end{equation}
where
\begin{eqnarray*}
\mathcal{X}&=&2(1-b)H(\dot{H}-\frac{k}{a^2})(H^2+
\frac{k}{a^2})f_R+(2-b)H(\dot{H}-\frac{k}{a^2})^2f_R
\\\nonumber&+&(1-b)(H^2+\frac{k}{a^2})^2\mathcal{F}
\partial_t(\frac{f_R}{\mathcal{F}}).
\end{eqnarray*}
Thus the condition to satisfy the GSLT is equivalent to
$\mathcal{X}\geq0$. In flat FRW universe, the GSLT is valid with the
constraints $\partial_t(\frac{f_R}{\mathcal{F}})\geq0,~H>0$ and
$\dot{H}\geq0$. Also, $\mathcal{F}$ and $f_R$ are positive in order
to keep $E>0$. If $b=1$, i.e., temperature between outside and
inside the horizon remains the same then the GSLT is valid only if
\begin{equation}\label{34}
\mathcal{J}=\left(\dot{H}-\frac{k}{a^2}\right)^2\frac{f_R}{\mathcal{F}}\geq0.
\end{equation}
For Eq.(\ref{10}), the effective EoS is defined as
$\omega_{eff}=-1-{2(\dot{H}-\frac{k}{a^2})}/{3(H^2+\frac{k}{a^2})}$.
Here $\dot{H}<\frac{k}{a^2}$ corresponds to quintessence region
while $\dot{H}>\frac{k}{a^2}$ represents the phantom phase of the
universe. It follows that GSLT in $f(R,T)$ gravity is satisfied in
both phantom and non-phantom phases. This result is compatible with
\cite{37} according to which entropy may be positive even at the
phantom era. Bamba and Geng \cite{26,27} also shown that second law
of thermodynamics can be satisfied in $f(R)$ and $f(T)$ theories of
gravity.

\section{Redefining the Dark Components}

In previous section, we have seen that an additional entropy term
$d_{\jmath}S'_h$ is produced in laws of thermodynamics. This can be
considered as the result of non-equilibrium description of the field
equations. If we redefine the \emph{dark components} so that the
extra entropy production term is vanished, then such formulation is
referred as an equilibrium description. It has been seen so far that
the equilibrium description does exist in modified theories of
gravity \cite{26,27,29} and extra entropy production term can be
removed.

Here, we discuss whether the equilibrium description of $f(R,T)$
gravity can be anticipated. In fact, we may reduce the entropy
production term through this description but it cannot be wiped out
entirely. We redefine the energy density and pressure of \emph{dark
components}. The (00) and (11) components of the field equations can
be rewritten as
\begin{eqnarray}\label{35}
3\left(H^2+\frac{k}{a^2}\right)&=&8{\pi}G_{eff}({\rho}_m+\rho_d),
\\\label{36}-2\left(\dot{H}-\frac{k}{a^2}\right)&=&8{\pi}G_{eff}(\rho_m
+\rho_d+p_d),
\end{eqnarray}
where $G_{eff}=\left(G+\frac{f_{T}(R,T)}{8\pi}\right)$ is the
effective gravitational coupling, $\rho_d$ and $p_d$ are the energy
density and pressure of \emph{dark components} given by
\begin{eqnarray}\nonumber
\rho_d&=&\frac{1}{8{\pi}G\mathcal{F}}\left[\frac{1}{2}(Rf_{R}-f)-3H(\dot{R}f_{RR}
+\dot{T}f_{RT})+3(1-f_R)(H^2\right.\\\label{37}&+&\left.\frac{k}{a^2})\right],\\\nonumber
p_d&=&\frac{1}{8{\pi}G\mathcal{F}}\left[-\frac{1}{2}(Rf_{R}-f)+2H(\dot{R}f_{RR}
+\dot{T}f_{RT})+\ddot{R}f_{RR}+\dot{R}^2f_{RRR}\right.\\\label{38}&+&\left.2\dot{R}\dot{T}f_{RRT}
+\ddot{T}f_{RT}+\dot{T}^2f_{RTT}-(1-f_R)(2\dot{H}+3H^2+\frac{k}{a^2})\right].
\end{eqnarray}
The EoS parameter $\omega_d$ in this description turns out to be
\begin{eqnarray}\nonumber
\omega_d&=&-1+\{\ddot{R}f_{RR}+\dot{R}^2f_{RRR}+2\dot{R}\dot{T}f_{RRT}
+\ddot{T}f_{RT}+\dot{T}^2f_{RTT}-H(\dot{R}f_{RR}
\\\nonumber&+&\dot{T}f_{RT})-2(1-f_R)(\dot{H}-\frac{k}{a^2})\}/\{\frac{1}{2}(Rf_{R}-f)-3H(\dot{R}f_{RR}
+\dot{T}f_{RT})\\\label{38a}&+&3(1-f_R)(H^2+\frac{k}{a^2})\},
\end{eqnarray}
In this case the expression of total energy exchange is given by
\begin{eqnarray}\label{38b}
q_{tot}=\frac{3}{8{\pi}G}(H^2+\frac{k}{a^2})\partial_t
\left(\frac{1}{\mathcal{F}}\right).
\end{eqnarray}
Since $\partial_t(f_T(R,T))\neq0$ in $f(R,T)$ gravity, so that
$q_{tot}$ does not vanish. So, we may not establish the equilibrium
picture of thermodynamics in this modified gravity. Hence, again we
need to consider the non-equilibrium treatment of thermodynamics.
This result differ from other modified gravitational theories due to
the matter dependence of the Lagrangian density. In $f(R)$ gravity
the redefinition of dark components result in local conservation of
energy momentum tensor of dark components \cite{26}. It is clear
from Eqs.(\ref{16a}) and (\ref{38a}) that the EoS parameter of
\emph{dark components} is not unique in both cases. Thus, one should
consider both formulations of the field equations in cosmic
discussions.

Now we check the validity of the first and second laws of
thermodynamics in this scenario.

\subsection{First Law of Thermodynamics}

In this representation of the field equations, the time derivative
of radius $\tilde{r}_A$ at the apparent horizon is given by
\begin{equation}\label{39}
d\tilde{r}_A=4{\pi}\tilde{r}^3_A{G\mathcal{F}}(\rho_{tot}+p_{tot})Hdt.
\end{equation}
Since in $f(R,T)$ gravity, the equilibrium description is not
feasible as it can be seen in modified gravitational theories such
that $f(R),~f(T)$ and scalar tensor gravity etc. Thus, we use the
Wald entropy relation $S_h=A/(4G_{eff})$ rather than introducing
Bekenstein-Hawking entropy. Using Eq.(\ref{39}), the horizon entropy
becomes
\begin{equation}\label{40}
\frac{1}{2{\pi}\tilde{r}_A}dS_h=4{\pi}\tilde{r}^3_A(\rho_{tot}
+p_{tot})Hdt+\frac{\tilde{r}_A}{2G}d\left(\frac{1}{\mathcal{F}}\right).
\end{equation}
The associated temperature of the apparent horizon is
\begin{equation}\label{41}
T_h=\frac{1}{2\pi\tilde{r}_A}
(1-\frac{\dot{\tilde{r}}_A}{2H\tilde{r}_A}).
\end{equation}
Equations (\ref{40}) and (\ref{41}) imply that
\begin{eqnarray}\label{42}
T_hdS_h=4{\pi}\tilde{r}^3_A(\rho_{tot}+p_{tot})Hdt-2{\pi}\tilde{r}^2
_A(\rho_{tot}+p_{tot})d\tilde{r}_A+\frac{{\pi}\tilde{r}^2_AT_h}{G}d\left
(\frac{1}{\mathcal{F}}\right).
\end{eqnarray}
Introducing the Misner-Sharp energy
\begin{equation}\label{43}
E=\frac{\tilde{r}_A}{2G\mathcal{F}}=V\rho_{tot},
\end{equation}
we obtain
\begin{equation}\label{44}
dE=-4{\pi}\tilde{r}^3_A(\rho_{tot}+p_{tot})Hdt+4{\pi}
\tilde{r}^2_A\rho_{tot}d\tilde{r}_A+\frac{\tilde{r}_A}{2G}d\left
(\frac{1}{\mathcal{F}}\right).
\end{equation}

The total work density is defined as \cite{35}
\begin{equation}\label{45}
W=-\frac{1}{2}T^{(tot)\alpha\beta}h_{\alpha\beta}=\frac{1}{2}
(\rho_{tot}-p_{tot}),
\end{equation}
By combining Eqs.(\ref{42}), (\ref{44}) and (\ref{45}), we obtain
the following expression of first law of thermodynamics
\begin{equation}\label{46}
T_hdS_h+T_hd_{\jmath}S_h=-dE+WdV,
\end{equation}
where
\begin{eqnarray}\nonumber
d_{\jmath}S_h&=&-\frac{\tilde{r}_A}{2T_hG}(1+2{\pi}\tilde{r}_AT_h)
d\left(\frac{1}{\mathcal{F}}\right)=-\mathcal{F}\left(\frac{E}{T_h}+S
_h\right)d\left(\frac{1}{\mathcal{F}}\right)\\\label{47}&=&-\frac{{\pi}
(4H^2+\dot{H}+3k/a^2)}{G(H^2+k/a^2)(2H^2+
\dot{H}+k/a^2)}d\left(\frac{1}{\mathcal{F}}\right)
\end{eqnarray}
is the additional term of entropy produced due to the matter
contents of the universe. It involves derivative of $f(R,T)$ with
respect to the trace of the energy-momentum tensor. Notice that the
first law of thermodynamics $T_hdS_h=-dE+WdV$ holds at the apparent
horizon of FRW universe in equilibrium description of modified
theories of gravity \cite{26,27,29}. However, in $f(R,T)$ gravity,
this law does not hold due to the presence of an additional term
$d_{\jmath}S_h$. This term vanishes if we take $f(R,T)=f(R)$ which
leads to the equilibrium description of thermodynamics in $f(R)$
gravity.

\subsection{Generalized Second Law of Thermodynamics}

To establish the GSLT in this formulation of $f(R,T)$ gravity, we
consider the Gibbs equation in terms of all matter field and energy
contents
\begin{equation}\label{48}
T_{tot}dS_{tot}=d(\rho_{tot}V)+p_{tot}dV,
\end{equation}
where $T_{tot}$ denotes the temperature of total energy inside the
horizon and $S_{tot}$ is the entropy of all the matter and energy
sources inside the horizon. In this case, the GSLT can be expressed
as
\begin{equation}\label{49}
\dot{S}_h+d_\jmath\dot{S}_h+\dot{S}_{tot}\geq0
\end{equation}
which implies that
\begin{equation}\label{50}
\frac{12\pi\mathcal{Y}}{bRG\mathcal{F}(H^2+\frac{k}{a^2})^2}\geq0,
\end{equation}
where
\begin{eqnarray*}
\mathcal{Y}&=&2(1-b)H(\dot{H}-\frac{k}{a^2})(H^2+
\frac{k}{a^2})+(2-b)H(\dot{H}-\frac{k}{a^2})^2
\\\nonumber&+&(1-b)(H^2+\frac{k}{a^2})^2\mathcal{F}
\partial_t(\frac{1}{\mathcal{F}}).
\end{eqnarray*}
Thus the GSLT is satisfied only if $\mathcal{Y}\geq0$. In case of
flat FRW universe, the GSLT is met with the conditions
$\partial_t(\frac{1}{\mathcal{F}})\geq0$, $H>0$ and $\dot{H}\geq0$.
In thermal equilibrium $b=1$, the above condition is reduced to the
following form
\begin{equation}\label{51}
\mathfrak{B}=\frac{12{\pi}H\left(\dot{H}-\frac{k}{a^2}\right)^2}
{G\left(H^2+\frac{k}{a^2}\right)^2R}\frac{1}{\mathcal{F}}\geq0,
\end{equation}
for $V=\frac{4}{3}{\pi}\tilde{r}^3_A$ and $R=6(\dot{H}+2H^2+k/a^2)$.
$\mathfrak{B}\geq0$ clearly holds when the Hubble parameter and
scalar curvature have same signatures. It can be seen that main
difference of results of $f(R,T)$ gravity with $f(R)$ gravity is the
term $\mathcal{F}=1+\frac{f_{T}(R,T)}{8{\pi}G}$. We remark that in
both definitions of \emph{dark components}, the GSLT is valid both
in phantom and non-phantom phases of the universe.
\section{Concluding Remarks}

The fact, $f(R,T)$ gravity is the generalization of $f(R)$ gravity
is based on coupling between matter and geometry \cite{30}. This
theory can be applied to explore several issues of current interest
in cosmology and astrophysics. We have discussed the laws of
thermodynamic at the apparent horizon of FRW spacetime in this
modified gravity. Akbar and Cai \cite{23} have shown that the
Friedmann equations for $f(R)$ gravity can be written into a form of
the first law of thermodynamics, $dE=TdS+WdV+Td\bar{S}$, where
$d\bar{S}$ is the additional entropy term due to non-equilibrium
thermodynamics. Bamba and Geng \cite{26,27} established the first
and second laws of thermodynamics at the apparent horizon of FRW
universe with both non-equilibrium and equilibrium descriptions.

We have found that the picture of equilibrium thermodynamics is not
feasible in $f(R,T)$ gravity even if we specify the energy density
and pressure of \emph{dark components} (see Section \textbf{4}).
Thus the non-equilibrium treatment is used to study the laws of
thermodynamics in both forms of the energy-momentum tensor of
\emph{dark components}. In $f(R,T)$ gravity, $q_{tot}$ does not
vanish so there exists some energy exchange with the horizon. The
non-equilibrium description can be interpreted as due to some energy
flow between inside and outside the apparent horizon. The first law
of thermodynamics is obtained at the apparent horizon in FRW
background for $f(R,T)$ gravity.

We observe that the additional entropy term is produced as compared
to GR, Gauss-Bonnet gravity \cite{18}, Lovelock gravity \cite{19,20}
and braneworld gravity \cite{21}. The equilibrium and
non-equilibrium description of thermodynamics in $f(R)$ gravity can
be obtained if the term $f_T(R,T)$ vanishes \emph{i.e.,} Lagrangian
density depends only on geometric part. We have established the GSLT
with the assumption that the total temperature inside the horizon
$T_{tot}$ is proportional to the temperature of the apparent horizon
$T_h$ and evaluated its validity conditions. The GSLT in $f(R)$
gravity follows from condition (\ref{33}) if $\mathcal{F}=1$. It is
concluded that in thermal equilibrium, GSLT is satisfied in both
phantom and non-phantom phases.


\vspace{.5cm}

\begin{thebibliography}{36}

\bibitem{1}Bennett, C.L. et al.: Astrophys. J. Suppl.
\textbf{148}(2003)1; Spergel, D.N. et al.:
Astrophys. J. Suppl. \textbf{148}(2003)175; Spergel, D.N. et al.:
Astrophys. J. Suppl. \textbf{170}(2007)377.

\bibitem{2}Perlmutter, S. et al.: Astrophys. J. \textbf{483}(1997)565;
Perlmutter, S. et al.: Nature \textbf{391}(1998)51; Perlmutter, S.
et al.: Astrophys. J. \textbf{517}(1999)565; Riess, A.G. et al.:
Astrophys. J. \textbf{607}(2004)665; Riess, A.G. et al.: Astrophys.
J. \textbf{659}(2007)98.

\bibitem{3}Hawkins, E. et al.: Mon. Not. Roy. Astron.
Soc. \textbf{346}(2003)78; Tegmark, M. et al.: Phys. Rev.
\textbf{D69}(2004)103501; Cole, S. et al.: Mon. Not. Roy. Astron.
Soc. \textbf{362}(2005)505.

\bibitem{4}Eisentein, D.J. et al.: Astrophys. J. \textbf{633}(2005)560.

\bibitem{5}Jain, B. and Taylor, A.: Phys. Rev. Lett. \textbf{91}(2003)141302.

\bibitem{6}Sahni, V. and Starobinsky, A.A.: Int. J. Mod. Phys.
\textbf{D9}(2000)373; Sahni, V.: Lect. Notes Phys.
\textbf{653}(2004)141; Padmanabhan, T.: Gen. Relativ. Gravit.
\textbf{40}(2008)529.

\bibitem{7}Caldwell, R.R.: Phys. Lett. \textbf{B545}(2002)23;
Nojiri, S. and Odintsov, S.D.: Phys. Lett. \textbf{B562}(2003)147;
ibid. \textbf{B565}(2003)1.

\bibitem{8}Feng, B., Wamg, X.L. and Zhang, X.M.:
Phys. Lett. \textbf{B607}(2005)35; Guo, Z.K. et al.: Phys. Lett.
\textbf{B608}(2005)177.

\bibitem{9}Kamenshchik, A., Moschella, U. and Pasquier, V.:
Phys. Lett. \textbf{B511}(2001)265; Bento, M.C., Bertolami, O. and
Sen, A.A.: Phys. Rev. \textbf{D66}(2002)043507.

\bibitem{10}Padmanabhan, T.: Phys. Rev. \textbf{D66}(2002)021301;
Sen, A.: JHEP \textbf{48}(2002)204.

\bibitem{11}Nojiri, S. and Odintsov, S.D.: Int. J. Geom. Methods Mod.
Phys. \textbf{4}(2007)115; Sotiriou, T. P. and S. Liberati.: Ann.
Phys. \textbf{322}(2007)935; Sotiriou, T.P and Faraoni, V.: Rev.
Mod. Phys. \textbf{82}(2010)451; De Felice, A. and Tsujikawa, S.:
Living Rev. Rel. \textbf{13}(2010)3.

\bibitem{12}Bardeen, J.M., Carter, B. and Hawking, S.W.: Commun.
Math. Phys. \textbf{31}(1973)161.

\bibitem{13}Hawking, S.W.: Commun. Math. Phys. \textbf{43}(1975)199.

\bibitem{14}Bekenstein, J.D.: Phys. Rev. \textbf{D7}(1973)2333.

\bibitem{15}Jacobson, T.: Phys. Rev. Lett. \textbf{75}(1995)1260.

\bibitem{16}Frolov, A.V. and Kofman, L.: JCAP \textbf{05}(2003)009.

\bibitem{17}Cai, R.G. and Kim, S.P.: JHEP \textbf{02}(2005)050.

\bibitem{18}Akbar, M. and Cai, R.G.: Phys. Rev. \textbf{D75}(2007)084003.

\bibitem{19}Cai, R.G. and Cao, L.M.: Phys. Rev. \textbf{D75}(2007)064008.

\bibitem{20}Cai, R.G., Cao, L.M., Hu, Y.P. and Kim, S.P.: Phys. Rev. \textbf{D78}(2008)124012.

\bibitem{21}Sheykhi, A., Wang, B. and Cai, R.G.: Nucl. Phys.
\textbf{B779}(2007)1; ibid. Phys. Rev. \textbf{D76}(2007)023515.

\bibitem{22}Eling, C., Guedens, R. and Jacobson, T.: Phys. Rev. Lett. \textbf{86}(2006)121301.

\bibitem{23}Akbar, M. and Cai, R.G.: Phys. Lett. \textbf{B648}(2007)243.

\bibitem{24}Gong, Y. and Wang, A.: Phys. Rev. Lett. \textbf{99}(2007)211301.

\bibitem{25}Bamba, K. and Geng, C.Q.: Phys. Lett. \textbf{B679}(2009)282.

\bibitem{26}Bamba, K. and Geng, C.Q.: JCAP \textbf{06}(2010)014.

\bibitem{27}Bamba, K. and Geng, C.Q.: JCAP \textbf{11}(2011)008.

\bibitem{28}Wu, S.-F., Wang, B., Yang, G.-H. and Zhang, P.-M.: Class.
Quantum Grav. \textbf{25}(2008)235018.

\bibitem{29}Bamba, K., Geng, C.Q. and Tsujikawa, S.:
Phys. Lett. \textbf{B668}(2010)101.

\bibitem{30}Harko, T., Lobo, F.S.N., Nojiri, S. and Odintsov, S.D.:
Phys. Rev. \textbf{D84}(2011)024020.

\bibitem{31}Houndjo, M.J.S.: Int. J. Mod. Phys.
\textbf{D}(to appear, 2012), arXiv:1107.3887.

\bibitem{31a}Gong, Y. and Wang, A.: Phys. Rev. Lett. \textbf{99}(2007)211301;
Eling, C.: JHEP \textbf{11}(2008)048; Elizalde, E. and Silva, P.J.:
Phys. Rev. \textbf{D78}(2008)061501; Wu, S.-F., Ge, X.-H., Zhang,
P.-M. and Yang, G.-H.: Phys. Rev. \textbf{D81}(2010)044034.

\bibitem{32}Landau, L.D. and Lifshitz E.M.: \emph{The Classical Theory
of Fileds} (Butterworth-Heinemann, 2002).

\bibitem{33}Wald, R.M.: Phys. Rev. \textbf{D48}(1993)3427.

\bibitem{33a}Brustein, R., Gorbonos, D. and Hadad, M.: Phys. Rev. \textbf{D79}(2009)044025.

\bibitem{34}Misner, C.W. and Sharp, D.H.: Phys. Rev. \textbf{136}(1964) B571;
Bak, D. and Rey, S.-J.: Class. Quantum Grav. \textbf{17}(2000)L83.

\bibitem{35}Hayward, S.A.: Class. Quantum Grav. \textbf{15}(1998)3147;
Hayward, S.A., Mukohyama, S. and Ashworth, M.: Phys. Lett.
\textbf{A256}(1999)347.

\bibitem{36}Izquierdo, G. and Pavon, D.: Phys. Lett. \textbf{B633}(2006)420.

\bibitem{37}Nojiri, S. and Odintsov, S.D.: Phys. Rev. \textbf{D72}(2005)023003.


\end{thebibliography}
\end{document}